\documentclass[twocolumn]{aastex702}

\shorttitle{GX~339--4: QPOs vs.\ BBN}
\shortauthors{Qin et al.}

\graphicspath{{./}{fig/}}

\usepackage{CJK}
\usepackage{txfonts}
\usepackage{mathrsfs}
\usepackage{enumitem}
\setlist{leftmargin=1.5em}

\usepackage{siunitx}
\hypersetup{%
  pdfstartview      = FitH,
  bookmarksnumbered = true,
  bookmarksopen     = true,
  pdftitle          = {GX 339-4: QPOs vs. BBN},
  pdfauthor         = {Qin et al.},
}

\newcommand{\subrm}[1]{_{\mathrm{#1}}}
\newcommand{\eqref}[1]{Equation~(\ref{#1})}
\renewcommand{\phs}{\phantom{-}}
\def\nbhyphen{\mbox{-}}
\def\nbendash{\mbox{--}}

\begin{document}
\begin{CJK*}{UTF8}{gbsn}

\title{Phase-resolved QPO Analysis of GX~339--4: Improved Technique and Consistent Behaviors between QPOs and Broadband Noise}

\author{Jin Qin (覃津)}
\email{qinj23@mails.tsinghua.edu.cn}
\affiliation{Department of Astronomy, Tsinghua University, Beijing 100084, People's Republic of China}

\correspondingauthor{Hua Feng}
\author{Hua Feng}
\email[show]{hfeng@ihep.ac.cn}
\affiliation{State Key Laboratory of Particle Astrophysics, Institute of High Energy Physics, Chinese Academy of Sciences, 19B Yuquan Road, Beijing 100049, People's Republic of China}

\author{Liang Zhang}
\email{zhangliang@ihep.ac.cn}
\affiliation{State Key Laboratory of Particle Astrophysics, Institute of High Energy Physics, Chinese Academy of Sciences, 19B Yuquan Road, Beijing 100049, People's Republic of China}

\author{Lian Tao}
\email{taolian@ihep.ac.cn}
\affiliation{State Key Laboratory of Particle Astrophysics, Institute of High Energy Physics, Chinese Academy of Sciences, 19B Yuquan Road, Beijing 100049, People's Republic of China}

\author{Qing-Cang Shui}
\email{shuiqc@ihep.ac.cn}
\affiliation{State Key Laboratory of Particle Astrophysics, Institute of High Energy Physics, Chinese Academy of Sciences, 19B Yuquan Road, Beijing 100049, People's Republic of China}

\author{Qing-Chang Zhao}
\email{zhaoqc@ihep.ac.cn}
\affiliation{State Key Laboratory of Particle Astrophysics, Institute of High Energy Physics, Chinese Academy of Sciences, 19B Yuquan Road, Beijing 100049, People's Republic of China}
\affiliation{University of Chinese Academy of Sciences, Chinese Academy of Sciences, 100049 Beijing, People's Republic of China}

\author{Shu Zhang}
\email{szhang@ihep.ac.cn}
\affiliation{State Key Laboratory of Particle Astrophysics, Institute of High Energy Physics, Chinese Academy of Sciences, 19B Yuquan Road, Beijing 100049, People's Republic of China}

\author{Shuang-Nan Zhang}
\email{zhangsn@ihep.ac.cn}
\affiliation{State Key Laboratory of Particle Astrophysics, Institute of High Energy Physics, Chinese Academy of Sciences, 19B Yuquan Road, Beijing 100049, People's Republic of China}
\affiliation{University of Chinese Academy of Sciences, Chinese Academy of Sciences, 100049 Beijing, People's Republic of China}

\begin{abstract} 
The nature of low-frequency quasi-periodic oscillations (QPOs) in black hole X-ray binaries remains unclear, and their relationship with the accompanying broadband noise (BBN) is still under debate.
Here, we propose an improved variational mode decomposition (VMD) technique.
Compared with the original algorithm that requires iterative, case-by-case parameter tuning, the new algorithm automatically and consistently determines the relevant VMD parameters based on the QPO central frequency and width measured from the power spectral density (PSD). 
This enables a more robust phase determination for QPOs and can also be applied to the study of BBN.
We found that, for low-frequency type-C QPOs without significant harmonics in the black hole X-ray binary GX~339--4, the spectral properties of the QPOs and BBN are statistically consistent with each other:
(1)~the photon index is positively correlated with count rate as a function of phase, and (2)~the PSD ratio spectra across different energy bands show no statistically significant QPO-like structures near the QPO frequencies, indicating that both components share the same energy dependence.
These suggest that QPOs and BBN may be driven by the same physical processes.
The QPO models based on geometric modulation struggle to account for the results, while those invoking corona oscillations are favored. 
\end{abstract}


\section{Introduction}

Transient black hole X-ray binaries (BHXRBs) transition through different spectral states during outbursts \citep[e.g.,][]{2006csxs.book..157M, 2010LNP...794...53B}.
In a typical full outburst, the source begins in the Hard State (HS), where the spectrum is dominated by a Comptonized component \citep{2006csxs.book..157M}. 
As the outburst progresses, it evolves through the Hard-Intermediate State (HIMS) and Soft-Intermediate State into the Soft State, which is dominated by thermal disk emission \citep{2006csxs.book..157M}. 
Finally, the source returns to the HS before fading into quiescence, completing a q-shaped hysteresis loop in the hardness--intensity diagram \citep[e.g.,][]{2001ApJS..132..377H,2004MNRAS.355.1105F,2005A&A...440..207B,2025ApJ...985..217H}.

BHXRBs in the HS and HIMS are known for high time variability.
Strong broadband noise (BBN) can be seen in the power spectral density (PSD) with a fractional rms variability of ${\sim}10\% \textendash 30\%$ \citep[e.g.,][]{2011BASI...39..409B}, and is argued to be a result of accretion rate variations propagating in the accretion flow \citep{1997MNRAS.292..679L,2011MNRAS.415.2323I}.
Type-C quasi-periodic oscillations (QPOs) manifesting as a narrow peak in the PSD are also seen in the bright HS and HIMS \citep{2002ApJ...564..962R,2005ApJ...629..403C,2019NewAR..8501524I}.
During the rising phase of an outburst, the central frequency of type-C QPOs evolves from ${\sim}\qty{.1}{Hz}$ to ${\sim}\qty{10}{Hz}$, and tightly correlates with the spectral index \citep{2003A&A...397..729V} and break frequency\footnote{Defined as $f\subrm{b}=\sqrt{f\subrm{c}^2+\Delta^2}$ \citep{1997A&A...322..857B}, where $f\subrm{c}$ and $\Delta$ are the central frequency and half width at half maximum of the Lorentzian component to describe the BBN.} of the BBN \citep{1999ApJ...514..939W}.
Harmonics and sub-harmonics are sometimes seen in the PSD with different rms--energy relations \citep[e.g.,][]{1997A&A...322..857B,2010ApJ...714.1065R,2016MNRAS.458.1778A} and lags \citep{2017MNRAS.464.2643V}.
Other types of QPOs are also observed but are less common.
Type-A and type-B QPOs are broader and occur in soft-intermediate state with weak red noise (rms $\lesssim 10\%$) \citep{1999ApJ...526L..33W,2005ApJ...629..403C,2019NewAR..8501524I}.
High-frequency QPOs (${\gtrsim}\qty{30}{Hz}$) are also observed in some sources \citep{1997ApJ...482..993M,2010LNP...794...53B}.

X-ray QPOs provide a powerful diagnostic for probing the physics of the inner accretion flow.
If QPOs are linked to general relativistic effects, they can be used to constrain black hole mass and spin \citep{2014MNRAS.437.2554M,2017MNRAS.467..145F}.
The latter remains notoriously difficult to measure and is often subject to significant systematic uncertainties \citep{2026NewAR.10201746Z}.
However, these constraints depend heavily on the underlying QPO model. Therefore, pinpointing the mechanisms behind QPOs is crucial for establishing them as an accurate diagnostic tool.

Although various QPO models have been proposed \citep[for a review, see][]{2019NewAR..8501524I}, there still is no consensus about the physical mechanism of the most commonly seen type-C QPOs; none of them can satisfactorily explain all the observed phenomena.
A prominent model for type-C QPOs attributes their origin to the Lense--Thirring precession of the hot inner flow \citep{2009MNRAS.397L.101I}.
In this framework, the accretion flow consists of an outer truncated thin disk and a hot inner flow that is misaligned with the black hole spin.
The misalignment causes the inner flow to precess around the spin axis, driven by the strong gravitational effects of the rotating black hole.
\citet{2009MNRAS.397L.101I} assumed that the whole hot inner flow precesses like a solid body at the angular momentum--weighted Lense--Thirring frequency.
Consequently, fluctuations in accretion can alter the radial distribution of the angular momentum, giving rise to the observed quasi-periodicity.
The quasi-periodic flux modulation results from a combination of the varying projected area, self-obscuration, and the supply of seed photons intercepted by the hot inner flow.
Furthermore, by assuming the BBN originates within the same precessing hot inner flow, \citet{2011MNRAS.415.2323I} successfully reproduced the strong correlation between the QPO frequency and break frequency of BBN. 

While the precession model succeeds in many respects, several observed phenomena remain difficult to explain.
It cannot explain the sub-harmonics of QPOs \citep{2019NewAR..8501524I} and the significantly softer second harmonic \citep{2016MNRAS.458.1778A}.
It is difficult to reconcile the predicted X-ray polarization as a function of QPO phase with the observed results, unless the system is viewed nearly edge-on \citep{2024ApJ...961L..42Z}.
The inferred truncation radius of the outer disk fails to reproduce the observed QPO frequency seen in GRS~1915+105 \citep{2022MNRAS.511..255N}.
More critically, \citet{2021ApJ...906..106M} argued that solid-body precession is unlikely to occur in the luminous HS, as the required (super)sonic accretion speeds force the flow into a diffusive regime where bending waves are damped and the disk aligns with the black hole spin at large radii. 

QPO phase-resolved spectroscopy has provided us an effective means to study the physical origin of QPOs \citep[e.g.,][]{2001ApJ...548..401T,2015MNRAS.446.3516I,2023ApJ...957...84S}. It allows us to explore how the spectral parameters vary with flux within the QPO cycle.
Existing techniques either require specific strong requirements, or needs to manually adjust empirical parameters for each observation.
These limitations hinder systematic analysis with large dataset.

GX~339--4 is an archetypal transient BHXRB characterized by rich and diverse QPO phenomena; extensive studies have significantly advanced our understanding of QPO physics \citep[e.g.,][]{2011MNRAS.418.2292M,2016MNRAS.458.1778A,2017ApJ...845..143Z,2019MNRAS.486.3451A,2021MNRAS.508..287S}.
In this work, we investigate the type\nbhyphen C QPOs and BBN in the 2007 and 2010 outbursts of GX~339\nbendash 4, using an improved phase-resolving technique based on the Hilbert--Huang transform. 
Observations of other outbursts are not included due to low statistics or unsuitable data modes (see \autoref{sec:obs-redn}).
In \autoref{sec:obs-redn}, we describe the sample selection and data reduction.
The analysis methods, including our improved algorithm, are described in \autoref{sec:methods}.
We show the results in \autoref{sec:results} and discuss their physical implications in \autoref{sec:disc}.

\section{Observations and Data Reduction} \label{sec:obs-redn}

We analyzed the archival data of GX~339--4 observed with the Proportional Counter Array \citep[PCA;][]{1996SPIE.2808...59J} onboard the \textit{Rossi X-ray Timing Explorer} (\textit{RXTE}) during the 2007 and 2010 outbursts.
The observations with type\nbhyphen C QPOs during the 1996 and 2004 outbursts have low rates ($\qtyrange{100}{300}{counts.s^{-1}.PCU^{-1}}$, compared with the selected ones in \autoref{tab:obs}).
The high time resolution data modes of the observations during the 2002 outburst have a poor spectral resolution (single bit modes with only 2--3~channels below $\qty{15}{keV}$).
As phase-resolved spectral analysis requires high statistics, good time resolution, and reasonable spectral resolution, we did not include the observations of these outbursts in our sample.
All observations selected for this study utilized the \verb|E_125us_64M_0_1s| data mode.

\begin{deluxetable*}{ccccccccccc}
\tablecaption{Selected observations and their starting time, background-corrected $\qtyrange{3.3}{20.2}{keV}$ PCU2 mean rate, exposure, best-fit QPO central frequency and HWHM, statistics of the PSD fitting, parameter $\alpha$ for VMD, and the fractional power of QPOs within the IMF as well as that of QPOs + BBN defined by \eqref{eq:power-ratio}.\label{tab:obs}}
\tablewidth{0pt}
\tablehead{%
\colhead{ObsID} &
\colhead{MJD Start} &
\colhead{Rate} &
\colhead{Exposure} &
\colhead{$f\subrm{QPO}$} &
\colhead{$\Delta\subrm{QPO}$} &
\colhead{$\chi^2$/dof} &
\colhead{$\alpha$} &
\colhead{$P'\subrm{QPO}/P'\subrm{tot}$} &
\colhead{$P'\subrm{src}/P'\subrm{tot}$} \\
&& \colhead{(counts s$^{-1}$)} & \colhead{(s)} & \colhead{(Hz)} & \colhead{($\qty{e-2}{Hz}$)} && \colhead{(s$^2$)}
}
\decimals
\startdata
92035-01-01-01 & 54128.944 & 651.1 &    \phn 3173 & $0.1416\pm0.0018$ & $0.67\pm0.29$ & 225/206 & 31.0 \phn & $57.5\%$ & $99.8\%$ \\
92035-01-01-03 & 54130.133 & 705.0 &    \phn 1287 & $0.1650\pm0.0028$ & $0.77\pm0.38$ & 165/161 & 23.2 \phn & $63.1\%$ & $99.8\%$ \\
92035-01-01-02 & 54131.105 & 742.9 &    \phn 3406 & $0.1751\pm0.0016$ & $0.63\pm0.25$ & 209/198 & 34.9 \phn & $60.3\%$ & $99.7\%$ \\
92035-01-01-04 & 54132.085 & 768.5 &    \phn 3375 & $0.2005\pm0.0016$ & $0.68\pm0.23$ & 205/192 & 30.3 \phn & $63.6\%$ & $99.7\%$ \\
92035-01-02-00 & 54133.002 & 796.2 &    \phn 3359 & $0.2320\pm0.0025$ & $1.47\pm0.39$ & 178/197 & \phn 6.38 & $51.9\%$ & $99.7\%$ \\
92035-01-02-01 & 54133.922 & 802.1 &    \phn 2972 & $0.2611\pm0.0023$ & $1.38\pm0.33$ & 175/200 & \phn 7.24 & $57.8\%$ & $99.7\%$ \\
92035-01-02-02 & 54135.033 & 817.0 &    \phn 3310 & $0.2917\pm0.0030$ & $2.41\pm0.47$ & 209/194 & \phn 2.31 & $62.5\%$ & $99.4\%$ \\
92035-01-02-03 & 54136.016 & 842.0 &    \phn 3293 & $0.3573\pm0.0029$ & $2.02\pm0.45$ & 236/211 & \phn 3.37 & $59.8\%$ & $99.3\%$ \\
92035-01-02-04 & 54136.998 & 906.1 &    \phn 3295 & $0.4279\pm0.0023$ & $1.63\pm0.33$ & 228/203 & \phn 5.24 & $66.8\%$ & $99.6\%$ \\
92035-01-02-08 & 54137.852 & 926.4 & \phn\phn 919 & $0.5361\pm0.0058$ & $2.91\pm0.97$ & 119/156 & \phn 1.62 & $61.6\%$ & $99.3\%$ \\
92035-01-02-07 & 54138.831 & 940.9 &    \phn 2680 & $0.8953\pm0.0031$ & $3.01\pm0.44$ & 226/215 & \phn 1.54 & $74.7\%$ & $99.2\%$ \\
\hline
95409-01-12-01 & 55283.144 & 560.4 &        15145 & $0.1630\pm0.0012$ & $1.22\pm0.19$ & 296/302 & \phn 9.14 & $44.8\%$ & $99.4\%$ \\
95409-01-13-03 & 55288.367 & 627.7 &    \phn 2029 & $0.2252\pm0.0016$ & $0.62\pm0.19$ & 144/166 & 36.4 \phn & $72.7\%$ & $99.6\%$ \\
95409-01-13-04 & 55290.722 & 661.4 &    \phn 2938 & $0.2879\pm0.0030$ & $1.32\pm0.47$ & 257/200 & \phn 7.90 & $59.3\%$ & $99.4\%$ \\
95409-01-13-02 & 55291.649 & 678.9 &    \phn 1728 & $0.3189\pm0.0048$ & $2.15\pm0.68$ & 151/169 & \phn 2.96 & $56.1\%$ & $99.2\%$ \\
95409-01-13-05 & 55292.778 & 710.7 & \phn\phn 736 & $0.3734\pm0.0054$ & $2.62\pm0.98$ & 124/130 & \phn 1.99 & $78.8\%$ & $99.1\%$ \\
95409-01-13-06 & 55294.124 & 726.3 & \phn\phn 768 & $0.4455\pm0.0055$ & $2.39\pm0.94$ & 153/142 & \phn 2.42 & $71.1\%$ & $99.4\%$ \\
95409-01-14-00 & 55295.001 & 708.8 & \phn\phn 992 & $0.5462\pm0.0052$ & $2.41\pm0.70$ & 122/147 & \phn 2.38 & $71.4\%$ & $98.9\%$ \\
95409-01-14-01 & 55296.249 & 745.1 & \phn\phn 656 & $1.0538\pm0.0062$ & $3.67\pm0.82$ & 150/159 & \phn 1.04 & $78.5\%$ & $99.0\%$ \\
\enddata
\end{deluxetable*}

We used \verb|heasoft| version~6.33.2 for data reduction.
Good time intervals are defined based on pointing offsets smaller than $\ang{.02}$, elevation angles larger than $\ang{10}$, and operation of at least two proportional counter units (PCUs) including PCU2.

\begin{figure}
\centering
\includegraphics[width=\columnwidth]{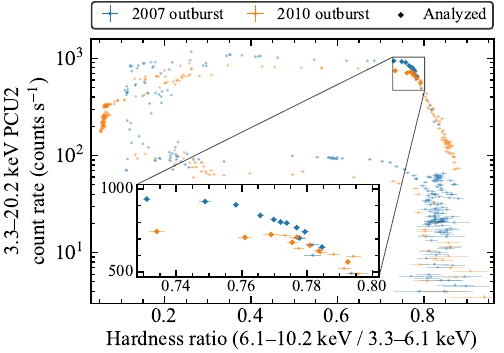}
\caption{Hardness--intensity diagrams of the 2007 and 2010 outbursts of GX~339--4. The lighted dots represent all \textit{RXTE} observations, and the squares represent the observations used for QPO analysis.}
\label{fig:hid}
\end{figure}

We extracted light curves from the filtered event data with a time resolution of $\qty[parse-numbers=false]{2^{-9}}{s} \approx \qty{2}{ms}$ from the absolute channels 5--49 ($\qtyrange{2.1}{20.6}{keV}$). 
We calculated the PSDs from the light curves with segment sizes of 128 and $\qty{256}{s}$, and selected those with a sufficient sampling for measuring the QPO components. 
The PSD is fitted with a model consisting of a constant component for the Poisson noise, 1--3 broad Lorentzians for the BBN, and a narrow Lorentzian ($Q=f\subrm{QPO}/2\Delta>2$) for each QPO component.
Only observations with a significant QPO detection (the Lorentzian normalization satisfies $N/\sigma_{N}>3$) are used for further analysis.
Further filtering to ensure the reliability of the phase-resolved results is also applied and will be detailed in \autoref{sec:improved-algo}.
To conclude, we selected 19 observations (see \autoref{tab:obs} and \autoref{fig:hid}) when the source was in the rising HS and early HIMS.
The PSDs of these observations contain one single significant type-C QPO without significant harmonics, and the frequencies of the QPOs range between ${\sim}\qtyrange{.1}{1}{Hz}$.
Since the second harmonics of GX~339--4 may have different physical origins indicated by the significantly softer spectra \citep{2016MNRAS.458.1778A}, and there are ambiguities in the identification of fundamental \citep{2019NewAR..8501524I}, we only focus on the observations without significant harmonics.
Observations in the late HIMS where significant harmonics are present will be analyzed and reported separately. 

We extracted the PCU2 spectra and generated the corresponding background and response files.
The responses were generated with \verb|pcarsp| in \verb|heasoft| version~6.31.1, since the newer version (6.33.2) of \verb|pcarsp| cannot produce the correct responses for the spectra extracted in gain-corrected modes.
The background is generated with \verb|pcabackest| using the bright background model.

The errors quoted in this work are all at the $1\sigma$ confidence level.

\section{Analysis Methods} \label{sec:methods}

\subsection{QPO Phase-resolving Techniques} \label{sec:qpo-phase-resolved-tech}

Broadly speaking, QPO phase-resolving techniques can be classified into two categories.
The first category involves statistically reconstructing QPO waveforms across different energy bands, and then the energy spectra are derived at different QPO phases from energy-dependent waveforms.
The QPO waveforms can be derived through cross-correlation \citep{2016MNRAS.460.2796S} or measuring waveform parameters from short light curve segments \citep{2015MNRAS.446.3516I}.
The former requires that there is no other variability component around the QPO frequency, while the latter requires the presence of a strong harmonic with fixed phase difference and amplitude ratio between the fundamental and harmonic.
These strong requirements limit their applications.

The second category first extracts the QPO signal to determine instantaneous phases, and then groups photons into discrete phase bins for spectral analysis.
This can be achieved via band-pass filtering \citep{2001ApJ...548..401T} or Hilbert--Huang transform \citep{1998RSPSA.454..903H,2015ApJ...815...74S,2023ApJ...957...84S}.
Compared with the first category, the second is more direct in spectral reconstruction and is less model-dependent.

The Hilbert--Huang transform \citep{1998RSPSA.454..903H} consists of two steps.
Firstly, it decomposes a light curve $g(t)$ into $K$ amplitude- and frequency-modulated quasi-sinusoidal signals (intrinsic mode functions; IMFs) $u_i(t)$, i.e.,
\begin{equation}
g(t) = \sum_{i=0}^{K-1} u_i(t) + r(t),
\end{equation}
where $r(t)$ is the residual and $u_i(t)$ can be expressed as
\begin{equation}\label{eq:imf}
u_i(t) = A_i(t)\cos \left[\phi_i(t)\right],
\end{equation}
with the instantaneous phase
\begin{equation}
\phi_i(t) = \phi_{i,0} + 2\pi\int_{t_0}^t f_i(\tau) \, d\tau,
\end{equation}
where $f_i(t)$ is the instantaneous frequency.
There are various algorithms to derive $u_i(t)$, including the original empirical mode decomposition \citep{1998RSPSA.454..903H}, variational mode decomposition \citep[VMD;][]{2014ITSP...62..531D}, and variational mode extraction \citep{2018IJBHI..22.1059N}.
In \autoref{sec:improved-algo}, we will briefly describe VMD and present an improved algorithm based on it.

Secondly, the instantaneous phases of the quasi-sinusoidal signals are derived as
\begin{equation}
\phi_i(t) = \mathop{\mathrm{atan2}}(\mathop{\mathscr{H}}(u_i(t)),u_i(t)),
\end{equation}
where the Hilbert transform $\mathop{\mathscr{H}}(u_i(t))$ shifts the phase of $u_i(t)$ by $\pi/2$ and is calculated as 
\begin{equation}
\mathop{\mathscr{H}}(u_i(t)) \equiv \frac{1}{\pi} \mathop{\rm p.v.} \int_{-\infty}^{+\infty} \frac{u_i(\tau)}{t-\tau} \, d\tau \,,
\end{equation}
where p.v.\ denotes the Cauchy principal value.
Thus, the Hilbert--Huang transform can be used to extract the QPO signal $u_i(t)$ and instantaneous phases $\phi_i(t)$, with which one can group the photons into discrete QPO phase bins to produce phase-resolved spectra.

\subsection{Algorithms for QPO Signal Extraction}\label{sec:improved-algo}

The VMD algorithm mentioned in \autoref{sec:qpo-phase-resolved-tech} fits the light curve $g(t)$ with quasi-sinusoidal signals $u_i(t)$ while balancing between two competing factors:\ minimizing the bandwidth of the IMFs in the frequency domain and maximizing the goodness of fit in the time domain.
Mathematically, this is achieved by minimizing the objective function \citep[Equation~(15) of][]{2014ITSP...62..531D}
\begin{eqnarray}
\mathop{\mathscr{L}}(\{u_i\}) &{}={}& 
\alpha \sum_{i=0}^{K-1} \left\Vert \frac{\partial}{\partial t} \left\{\left[ \left( \delta(t) + \frac{\mathrm{j}}{\pi t} \right) * u_i(t) \right] \mathrm{e}^{-\mathrm{j}2\pi \bar{f}_i t} \right\} \right\Vert^2 \nonumber\\
&& {} + \left\Vert g(t) - \sum_{i=0}^{K-1} u_i(t) \right\Vert^2, \label{eq:vmd-objf}
\end{eqnarray}
where $\bar{f}_i$ is the central frequency of the $i$th IMF, the first term measures the IMF bandwidth, and the second term is the residual.
The balance is governed by the parameter $\alpha$: the larger the value of $\alpha$, the narrower the bandwidth.

In practice, to solve \eqref{eq:vmd-objf}, one needs to assume an initial $u_i(t)$ and solve the equation iteratively. The IMF frequencies are randomly assigned or evenly distributed \citep{vmdpy}.
As a result, the QPOs may not be captured by any IMFs if $K$ is small.
Therefore, a larger $K$ is needed \citep{2023ApJ...957...84S}, though many IMFs are useless for our purpose.
The value of $\alpha$ is determined by comparing the shapes of the QPO and IMF in the PSD visually until they match \citep[see \autoref{fig:phase-resolved-results}(a) for an example;][]{2023ApJ...957...84S}.
The determination of $K$ and $\alpha$ needs iterations, which are time-consuming and experience-dependent. As the procedure needs to be done for each observation, it is hard to keep consistency among different observations.

To overcome these issues, we developed an algorithm\footnote{The implementation of the improved VMD algorithm is available at \url{https://github.com/ChynJin/vmd4qpo} and archived in Zenodo \citep{vmd4qpo}.}
that improves the efficiency and consistency via automatic parameterization by introducing
two prior constraints: the IMF central frequency is fixed at the best-fit QPO frequency, and the IMF should have a bandwidth similar to that of the QPOs.
Given the light curve, we first calculated and fit the PSD with multiple Lorentzians for QPOs and BBN and a constant for Poisson noise, and obtained the central frequency ($f_i$) and half width at half maximum (HWHM; $\Delta_i$) of the $i$th QPO component (also the $i$th QPO IMF).
An additional zero-frequency ($f_0 = 0$) IMF is needed to describe the low-frequency variations, with the HWHM $\Delta_0$ equal to the HWHM of the narrowest QPO.
We note that $\Delta_0$ does not affect the results significantly even with a decrease by an order of magnitude. 
With the frequencies known a priori, the Fourier transform of the $k$th IMF can be derived from \eqref{eq:vmd-objf} as
\begin{equation} \label{eq:analytic-vmd-diff-alpha}
\hat{u}_k(f) = \frac{\hat{g}(f)}{\left[8\pi^2+\sum_{i=0}^{K-1}\alpha_i^{-1}(f-f_i)^{-2}\right]\alpha_k(f-f_k)^2} \; ,
\end{equation}
where $\hat{g}(f)$ is the Fourier transform of the original light curve. 
We note that we did not require $\alpha$ to be the same in all IMFs as in the original VMD algorithm, which is inconsistent with observations \citep{2010ApJ...714.1065R}.

$\alpha$ can be directly computed, if we required that the IMF power drops rapidly at frequencies away from the central frequency (a narrow-band filter). 
In practice, we set ${P_k(f_k + 3\Delta_k) =\frac{1}{4} P(f_k + 3\Delta_k)}$, where $P_k(f) \propto |\hat{u}_k(f)|^2$ is the power of the $k$th IMF and $P(f) \propto |\hat{g}(f)|^2$ is the power of the original light curve.
We note that the coefficients used here (i.e., 1/4 and 3) have no significant impact on the result.
This leads to an analytical solution\footnote{We note that the Python library \texttt{vmdpy} \citep{vmdpy} used in \citet{2023ApJ...957...84S} uses a different definition of $\alpha$, with ${\alpha_{\tt vmdpy} = 2\alpha\left( \frac{2\pi}{\Delta t} \right)^2}$, where $\Delta t$ is the time resolution of the light curve.} of the set of $\alpha$ from the following linear equations about $\alpha_i^{-1}$:
\begin{equation} \label{eq:alpha}
\sum_{i=0}^{K-1}\frac{2\delta_{ik}-1}{(f_k-f_i+3\Delta_k)^2\alpha_i} = 8\pi^2,\; \text{for} \; k = 0,\ldots,K-1 \, ,
\end{equation}
where $\delta_{ik}=1$ (for $i=k$) or 0 (for $i\neq k$). 

In the case where $\alpha$ is the same for all IMFs or there is only one IMF of interest, simple expressions can be obtained as follows,
\begin{equation} \label{eq:analytic-vmd}
\hat{u}_k(f) = \frac{\hat{g}(f)}{\left[8\pi^2\alpha+\sum_{i=0}^{K-1}(f-f_i)^{-2}\right](f-f_k)^2} \; ,
\end{equation}
and
\begin{equation}
\alpha = \frac{1}{4\pi^2(3\Delta_k)^2} - \frac{1}{8\pi^2}\sum_{i=0}^{K-1} \frac{1}{(f_k-f_i+3\Delta_k)^2} \; .
\end{equation}

From \eqref{eq:analytic-vmd-diff-alpha}, one can see that VMD essentially serves as a band-pass filter, similar to the methods used in \citet{2001ApJ...548..401T} and \citet{2016MNRAS.458.3655V}.
Compared with those methods (Kaiser and optimal filters), VMD considers the broadening of the quasi-sinusoidal signal in a more physically motivated way (i.e., \eqref{eq:imf}).

In comparison with the original VMD, our improvements are in practical aspects.
With fixed QPO frequencies, \eqref{eq:vmd-objf} can be analytically solved without iterative optimization, which relies on the initialization of $u_i(t)$ and is time-consuming and less robust.
The best-fit HWHMs of the QPOs are used to determine a reasonable bandwidth controlling factor $\alpha$, which is manually chosen before.
Such automatic parameterization eliminates the trial-and-error process for each observation, and can keep consistency among different observations.

We emphasize that the difference between the two algorithms lies in the determination of $\alpha$ and $K$ --- both algorithms produce identical results if the same values are used, as both solve \eqref{eq:vmd-objf}.
However, these two parameters are selected empirically in the original algorithm and therefore depend on human experience, while the new algorithm makes the determination objective.
Without human intervention, the new algorithm is also more suitable for automated and large-scale processing.

We also emphasize the caveats with all these VMD-like techniques.
Poisson fluctuations can also be fitted by the IMFs and produce spurious count rate modulations
(see Appendix~\ref{appx:white-noise-modulation} for more details). 
To eliminate the effect, we always used PCU2 data for spectral analysis, and used the light curves summed over other PCU(s) for phase determination using VMD, because the white noise is not correlated between detectors.

To quantify how reliable the reconstructed phases are, we estimated the contribution from different power components to the QPO IMF:
\begin{equation}\label{eq:power-ratio}
\frac{P'\subrm{c}}{P'\subrm{tot}} = \int_{f_i-3\Delta_i}^{f_i+3\Delta_i} \eta(f) P\subrm{c}(f) \, d f  \,\bigg/\,  \int_{f_i-3\Delta_i}^{f_i+3\Delta_i} \eta(f) P\subrm{tot}(f) \, d f  \, ,
\end{equation}
where $\eta(f) \equiv \frac{P\subrm{IMF}(f)}{P\subrm{tot}(f)} = \frac{|\hat{u}_i(f)|^2}{|\hat{g}(f)|^2}$, $P\subrm{tot}(f)$ is the total power density, $P\subrm{IMF}(f)$ is the power density of the QPO IMF~$i$, subscript ``c'' represents the QPO, BBN, or source (the sum of QPO and BBN), and $P\subrm{c}(f)$ is the corresponding power density (the best-fit model components).
The larger $P'\subrm{src}/P'\subrm{tot}$ is, the less the reconstructed phases are affected by Poisson fluctuations.
We discarded observations where the contribution of Poisson noise dominates.
For all the observations used in this work, the contribution of Poisson noise to the IMF power is less than 2\%, and the QPO contribution varies between 45\%--80\% (\autoref{tab:obs}).
We also discarded time intervals where the IMF amplitude is among the lowest 10\%. We grouped the events into six phase bins and calculated phase-resolved spectra.

\subsection{Spectral Fitting}

We fit the spectra in $\qtyrange{3}{30}{keV}$ using \texttt{PyXspec} \citep{2021ascl.soft01014G}.
Channel~11 of the spectra is ignored since PCU2 contains no data in it for the gain-corrected mode \verb|E_125us_64M_0_1s| \citep{2016MNRAS.460.2796S}.
We adopted the spectral model \texttt{TBabs\allowbreak *\allowbreak edge\allowbreak (nthComp\allowbreak +\allowbreak diskbb\allowbreak +\allowbreak gaussian)} in \texttt{xspec}, where \texttt{diskbb} represents the standard disk emission and \texttt{nthComp} represents the Comptonization of disk photons. 
The broad iron line is modeled with a Gaussian component, without which significant residuals exist and the fit becomes unacceptable (typical reduced $\chi^2 \gtrsim 4$).
An iron edge is also included.
A systematic error of $0.5\%$ was added in the fits.

We first fit the time-averaged spectrum of each observation.
The hydrogen column density is fixed at ${N\subrm{H} = \qty{6e21}{cm^{-2}}}$ \citep{2020ApJ...899...44W} due to the lack of spectral coverage below $\qty{3}{keV}$.
We also tested $N\subrm{H}$ in the range of $(4\textendash8)\times\qty{e21}{cm^{-2}}$ and obtained consistent results for the continuum.
The electron temperature of \texttt{nthComp} is fixed at $\qty{40}{keV}$ as it consistently exceeds the energy band in the fit; we obtained well consistent results if we fixed it at any values in the range of $\qtyrange{20}{100}{keV}$.
The seed photon temperature of \texttt{nthComp} is linked to the inner disk temperature of \texttt{diskbb}.
The Gaussian energy cannot be reliably constrained, as there are no data in the $\qtyrange{6.1}{6.5}{keV}$ bin for PCU2 in the data mode that we used \citep{2016MNRAS.460.2796S}.
We therefore fixed the Gaussian central energy at $\qty{6.4}{keV}$ to align with the conclusion from previous observations \citep{2001ApJ...553..394F}, that the Fe~K$\alpha$ in GX~339--4 shifts from $\qty{6.4}{keV}$ at high fluxes to $\qty{6.7}{keV}$ at low fluxes.
We adopted $\sigma = \qty{1}{keV}$ because the line is known to be relatively broad \citep[e.g.,][]{2002MNRAS.332..856N}.
We also tried to fix the centroid energy at $\qty{6.7}{keV}$ and $\sigma$ at $\qty{2}{keV}$ and obtained well consistent results for the continuum components.
Therefore, the results presented here are insensitive to the iron line modeling.

We then fit the phase-resolved spectra in the same manner, using the time-averaged, best-fit values as the initial values.
The only exception is the \texttt{edge} component, whose parameters were fixed at the best-fit values derived from the time-averaged spectrum.
Due to a lack of coverage below $\qty{3}{keV}$, the errors of the parameters of \texttt{diskbb} are large. 
We therefore only discuss the modulations of the \texttt{nthComp} parameter, in particular the photon index $\Gamma$ in this work. 

\subsection{Comparison with BBN}

Type-C QPOs are always associated with strong BBN, and the power contribution from BBN to the QPO IMF cannot be neglected (${\sim}20\% \textendash 55\%$, see \autoref{tab:obs}).
As mentioned in \autoref{sec:improved-algo}, VMD actually fits the light curve with quasi-sinusoidal signals in a narrow frequency band. 
This means that it can also be applied on the BBN component if we fix the IMF to frequencies where the BBN dominates (see \autoref{fig:phase-resolved-results}a).
BBN has been argued to be due to fluctuation propagation, where the variability at different frequencies originates from different radii \citep{1997MNRAS.292..679L,2011MNRAS.415.2323I}.
We selected a series of central frequencies between ${\sim}0.5f_1$ and ${\sim}\qty{7}{Hz}$, where $f_1$ is the QPO frequency, to study the local BBN properties over the whole PSD where the power density is high.
We also required that the BBN IMF separates from the QPO IMF by at least six times the sum of their HWHMs, and the BBN contribution to the IMF power is larger than 50\%.
The width of the BBN IMF is set to have the same $Q$ factor as the QPO, which is motivated by the fact that the $Q$ factors of the fundamental and the second harmonic are similar \citep{2010ApJ...714.1065R}. 

In the presence of strong QPOs, the large power of QPOs may contaminate the BBN IMF near the QPO frequency.
To fully eliminate the contamination, we always included the QPO IMF when extracting the BBN IMF (i.e., three IMFs for the study of BBN: $f_0=\qty{0}{Hz}$, $f_1$ at QPOs, and $f_2$ for BBN).
This ensures that the power around the QPOs is absorbed by the QPO IMF and does not leak into the BBN IMF; mathematically, this is guaranteed by making the denominator of \eqref{eq:analytic-vmd-diff-alpha} infinity for the BBN IMF ($k=2$) at the QPO frequency $f=f_1$.

\section{Results} \label{sec:results}

\begin{figure*}
\centering
\includegraphics[width=\textwidth,page=9]{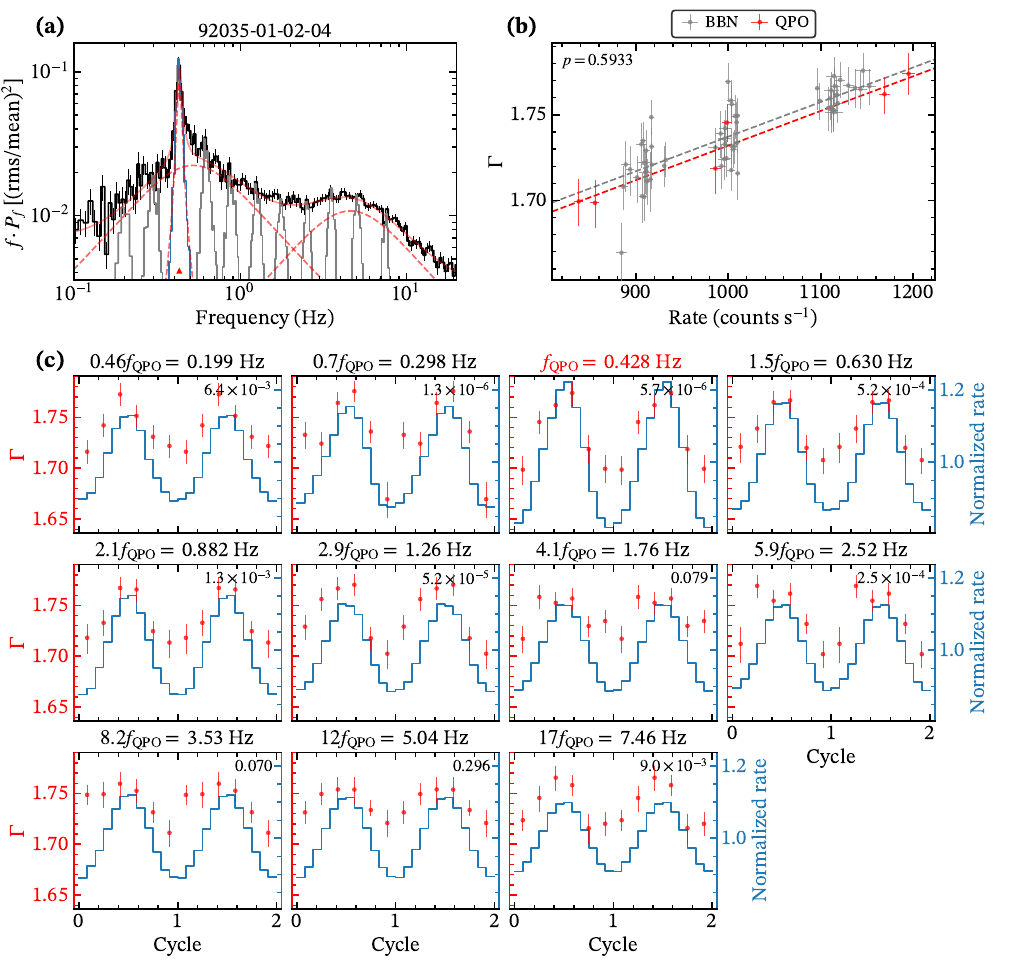}
\caption{PSD and phase-resolved photon index ($\Gamma$). 
(a)~PSD calculated from the observed light curve (black), with Lorentzian decompositions (red dashed) and the total model (red solid), QPO IMF (blue solid), and BBN IMFs (gray solid).
(b)~Photon index vs.\ mean rate of each phase bin of all IMFs (red for QPOs and gray for BBN). The dashed line represents the best-fit linear model to all BBN (gray) and QPO (red) data points. The $p$-value under the null hypothesis that the QPO and BBN IMFs follow the same relation is displayed. 
(c)~$\Gamma$ (red points) and normalized count rate (blue lines) as a function of IMF phase cycle. The $p$-value under the null hypothesis of a constant $\Gamma$ over phase is displayed in each panel. The complete figure set (19~images) is available in the online journal.}
\label{fig:phase-resolved-results}
\end{figure*}

In \autoref{fig:phase-resolved-results}(c), the phase modulation of photon index $\Gamma$ is shown, for both the QPO and BBN IMFs measured in one observation as an example.
The results of the other observations support the same conclusion to be discussed below, and are available in the online journal.
We can see that both the count rate and photon index vary in a nearly sinusoidal shape as a function of phase.
We calculated the $p$-values under the null hypothesis that there is no phase modulation of photon index, displayed in the top right corner of each panel.
For most of the IMFs, a small $p$ (${<}0.05$) is obtained, indicative of significant photon index variations over phases.
Large $p$-values found in a few cases are exclusively due to large errors.

\begin{deluxetable*}{ccccccccc}
\tablecaption{Best-fit parameters of the linear model $\Gamma = \kappa x + \beta$ for the $\Gamma$--rate relations for the QPO and BBN IMFs, respectively, and their differences. The last column lists the $p$-value from Chow test, which evaluates whether the two sets share the same $\Gamma$--rate relation.
\label{tab:fit-res}}
\tablewidth{0pt}
\tablehead{%
\colhead{} &
\multicolumn{3}{c}{Slope ($\qty{e-4}{counts^{-1}.s}$)} &
\colhead{} &
\multicolumn{3}{c}{Intercept} \\ \cline{2-4} \cline{6-8}
\colhead{ObsID} &
\colhead{$\kappa\subrm{QPO}$} &
\colhead{$\kappa\subrm{BBN}$} &
\colhead{$\kappa\subrm{QPO}-\kappa\subrm{BBN}$} &
\colhead{} &
\colhead{$\beta\subrm{QPO}$} &
\colhead{$\beta\subrm{BBN}$} &
\colhead{$\beta\subrm{QPO}-\beta\subrm{BBN}$} &
\colhead{$p$}
}
\decimals
\startdata
92035-01-01-01 & $2.07\pm0.33$ & $2.13\pm0.16$ & $    -0.06\pm0.36$ &  & $1.504\pm0.024$ & $1.502\pm0.012$ & $\phs 0.002\pm0.027$ & 0.9088 \\
92035-01-01-03 & $3.16\pm0.45$ & $2.36\pm0.18$ & $\phs 0.80\pm0.49$ &  & $1.415\pm0.038$ & $1.479\pm0.015$ & $    -0.064\pm0.041$ & 0.4630 \\
92035-01-01-02 & $1.97\pm0.34$ & $1.98\pm0.13$ & $    -0.01\pm0.37$ &  & $1.508\pm0.029$ & $1.510\pm0.011$ & $    -0.002\pm0.031$ & 0.8638 \\
92035-01-01-04 & $1.12\pm0.30$ & $1.43\pm0.13$ & $    -0.31\pm0.32$ &  & $1.580\pm0.025$ & $1.558\pm0.011$ & $\phs 0.022\pm0.027$ & 0.6381 \\
92035-01-02-00 & $1.50\pm0.22$ & $2.18\pm0.10$ & $    -0.68\pm0.24$ &  & $1.557\pm0.020$ & $1.497\pm0.009$ & $\phs 0.060\pm0.022$ & 0.1283 \\
92035-01-02-01 & $2.22\pm0.29$ & $2.09\pm0.12$ & $\phs 0.13\pm0.32$ &  & $1.489\pm0.027$ & $1.502\pm0.011$ & $    -0.013\pm0.029$ & 0.9557 \\
92035-01-02-02 & $2.46\pm0.25$ & $1.54\pm0.12$ & $\phs 0.92\pm0.28$ &  & $1.475\pm0.024$ & $1.562\pm0.011$ & $    -0.087\pm0.026$ & 0.0592 \\
92035-01-02-03 & $2.24\pm0.26$ & $1.94\pm0.12$ & $\phs 0.30\pm0.29$ &  & $1.496\pm0.025$ & $1.526\pm0.012$ & $    -0.030\pm0.028$ & 0.7662 \\
92035-01-02-04 & $2.02\pm0.28$ & $2.02\pm0.13$ & $\phs 0.00\pm0.30$ &  & $1.530\pm0.029$ & $1.535\pm0.013$ & $    -0.005\pm0.031$ & 0.5933 \\
92035-01-02-08 & $1.87\pm0.48$ & $1.24\pm0.22$ & $\phs 0.63\pm0.53$ &  & $1.559\pm0.051$ & $1.624\pm0.022$ & $    -0.066\pm0.056$ & 0.6979 \\
92035-01-02-07 & $2.68\pm0.32$ & $2.54\pm0.16$ & $\phs 0.15\pm0.35$ &  & $1.500\pm0.034$ & $1.520\pm0.017$ & $    -0.020\pm0.038$ & 0.7614 \\
\hline
95409-01-12-01 & $2.11\pm0.19$ & $2.08\pm0.08$ & $\phs 0.03\pm0.21$ &  & $1.526\pm0.012$ & $1.528\pm0.005$ & $    -0.002\pm0.013$ & 0.9873 \\
95409-01-13-03 & $1.37\pm0.47$ & $2.64\pm0.20$ & $    -1.27\pm0.51$ &  & $1.589\pm0.033$ & $1.500\pm0.014$ & $\phs 0.089\pm0.036$ & 0.1246 \\
95409-01-13-04 & $1.57\pm0.34$ & $2.47\pm0.14$ & $    -0.90\pm0.37$ &  & $1.578\pm0.025$ & $1.512\pm0.011$ & $\phs 0.066\pm0.028$ & 0.2245 \\
95409-01-13-02 & $2.63\pm0.36$ & $1.66\pm0.18$ & $\phs 0.97\pm0.40$ &  & $1.499\pm0.028$ & $1.575\pm0.014$ & $    -0.076\pm0.031$ & 0.3006 \\
95409-01-13-05 & $1.83\pm0.54$ & $0.98\pm0.31$ & $\phs 0.85\pm0.62$ &  & $1.553\pm0.042$ & $1.624\pm0.025$ & $    -0.071\pm0.049$ & 0.5837 \\
95409-01-13-06 & $2.79\pm0.55$ & $1.40\pm0.27$ & $\phs 1.38\pm0.62$ &  & $1.492\pm0.046$ & $1.609\pm0.022$ & $    -0.117\pm0.051$ & 0.2618 \\
95409-01-14-00 & $2.58\pm0.49$ & $2.04\pm0.25$ & $\phs 0.54\pm0.55$ &  & $1.534\pm0.041$ & $1.583\pm0.020$ & $    -0.049\pm0.045$ & 0.6972 \\
95409-01-14-01 & $3.00\pm0.61$ & $2.96\pm0.32$ & $\phs 0.04\pm0.69$ &  & $1.558\pm0.052$ & $1.563\pm0.027$ & $    -0.005\pm0.058$ & 0.9871 \\
\enddata
\end{deluxetable*}

To compare the relation of photon index versus count rate between the QPO and BBN IMFs, we plotted $\Gamma$ measured in individual phase bins from all IMFs as a function of the mean count rate of the corresponding phase bin in \autoref{fig:phase-resolved-results}(b).
A similar linear correlation is shown for both the QPO and BBN IMFs.
To check whether the linear relations are different for the QPO and BBN IMFs, we performed the Chow test \citep{chow-test} for each observation.
We respectively fit the data points for the QPO, BBN, and all IMFs with linear functions $\Gamma = \kappa x + \beta$, where $x$ is the count rate,
and calculated the residual sum of squares (RSS) for them, respectively:
\begin{equation}
\mathrm{RSS}\subrm{c} = \sum_{i \,\in\, \mathrm{c}} \frac{[\Gamma_i - (\kappa\subrm{c}x_i + \beta\subrm{c})]^2}{\sigma^2(\Gamma_i)},
\end{equation}
where ``c'' represents QPOs, BBN or all IMFs.
The best-fit slopes and intercepts for the QPO and BBN IMFs are listed in \autoref{tab:fit-res}.
Then the test statistic
\begin{equation}
F = \frac{[\mathrm{RSS}\subrm{all}-(\mathrm{RSS}\subrm{QPO}+\mathrm{RSS}\subrm{BBN})]/2}{(\mathrm{RSS}\subrm{QPO}+\mathrm{RSS}\subrm{BBN})/(N\subrm{all}-4)}
\end{equation}
follows $F$-distribution with 2 and $(N\subrm{all}-4)$ degrees of freedom, where $N\subrm{all}$ is the total number of data points.
We then calculated the $p$-value under the null hypothesis that QPO and BBN data points follow the same linear relation, and showed it in the top left corner of \autoref{fig:phase-resolved-results}(b), as well as \autoref{tab:fit-res}.
We obtained $p > 0.05$ in all observations, suggesting that there is no evidence that they follow different relations.

We caution that $p>0.05$ does not prove that the relations between the QPO and BBN IMFs are the same; rather, it merely indicates that, within the current uncertainties, there is no statistically significant evidence that they differ.
We listed the differences in slope and intercept between the QPO and BBN IMFs and their errors in \autoref{tab:fit-res}.
The differences are all below $3\sigma$ except for 92035-01-02-02, which shows a ${\sim}3.3\sigma$ difference for a single parameter. 
However, when comparing two linear relations, the two parameters should be considered jointly, since there is degeneracy between them in a way that a smaller slope leads to a larger intercept.
If one considers the $p$-value from Chow test, the chance probability is only ${\sim}0.06$ under the null hypothesis.
As we have 19~observations in the sample, this is consistent with statistical fluctuations.

\begin{deluxetable}{cccc}
\tablecaption{Average difference in slope and intercept for observations with different $P'\subrm{QPO}/P'\subrm{tot}$. The second column is the number of observations in each group. \label{tab:diff}}
\tablewidth{0pt}
\tablehead{%
\colhead{$P'\subrm{QPO}/P'\subrm{tot}$} &
\colhead{$N\subrm{obs}$} &
\colhead{$\overline{\kappa\subrm{QPO}-\kappa\subrm{BBN}}$} &
\colhead{$\overline{\beta\subrm{QPO}-\beta\subrm{BBN}}$}
}
\decimals
\startdata
$40\%\textendash 58\%$ & 5 & $    -0.06\pm0.28$ & $\phs 0.002\pm0.021$ \\
$58\%\textendash 63\%$ & 5 & $\phs 0.25\pm0.34$ & $    -0.018\pm0.030$ \\
$63\%\textendash 72\%$ & 5 & $\phs 0.19\pm0.41$ & $    -0.022\pm0.036$ \\
$72\%\textendash 80\%$ & 4 & $    -0.09\pm0.49$ & $\phs 0.011\pm0.043$ \\
\enddata
\end{deluxetable}

There are several factors that can bias our statistical analysis.
Firstly, the data points involved in the test may not be independent.
This will effectively reduce the true degrees of freedom and cause an overestimation of $F$ and consequently a smaller $p$-value.
Another effect is the contamination of the BBN to the QPO IMF (see the $P'\subrm{QPO}/P'\subrm{tot}$ in \autoref{tab:obs}).
If this is the case, the slope and intercept of the $\Gamma$--rate relation for QPOs will tend to be similar with those for BBN towards small $P'\subrm{QPO}/P'\subrm{tot}$.
We therefore grouped the observations by $P'\subrm{QPO}/P'\subrm{tot}$ and calculated the average difference in slope and intercept for each group, shown in \autoref{tab:diff}, and found no correlation.

\begin{figure}
\centering
\includegraphics[width=\columnwidth,page=9]{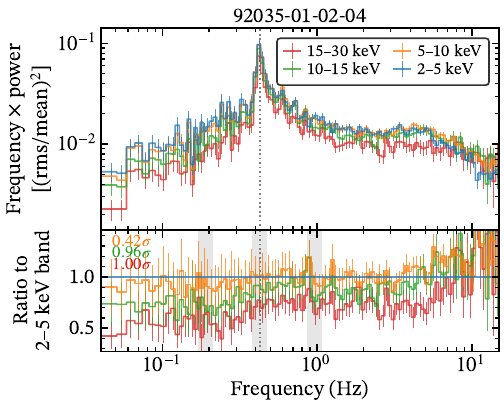}
\caption{PSDs in four energy bands (top) and their ratios against the one in $\qtyrange{2}{5}{keV}$ (bottom).
The significance of the difference between the PSD ratio measured in the QPO frequencies and that measured in the BBN frequencies for different energy bands is displayed in the lower panel.
The shaded regions indicate the frequency ranges to measure the PSD ratios.
The vertical dashed line marks the QPO central frequency. The complete figure set (19~images) is available in the online journal.}
\label{fig:psd-ratio}
\end{figure}

To further investigate if the the QPOs and BBN share the same spectro-timing behaviors, we extracted the PSDs in four different energy bands ($\numrange{2}{5}$, $\numrange{5}{10}$, $\numrange{10}{15}$, and $\qtyrange{15}{30}{keV}$) for each observation, and calculated the ratios against the PSD in the lowest band, shown in \autoref{fig:psd-ratio} for an example. 
The results of the other observations support the same conclusion to be discussed below, and are available in the online journal.
As one can see in \autoref{fig:psd-ratio}, the PSD ratios show a featureless transition over the QPO frequencies.

To quantify the significance of any possible structures in the PSD ratio spectrum around the QPO frequencies, we compared the PSD ratios 
around the frequencies dominated by QPOs and BBN, respectively.
Specifically, we compared the mean PSD ratio within $f\subrm{QPO}\pm 3\Delta$ (QPO region) with the geometric mean in $0.4f\subrm{QPO}\textendash 0.5f\subrm{QPO}$ and $2f\subrm{QPO}\textendash 2.5f\subrm{QPO}$ (BBN regions).
The significance of any difference (excess or deficit) between them is shown in \autoref{fig:psd-ratio} (the upper left corner of the bottom panel) at different energy bands, and is all less than $2\sigma$ for all observations, including 92035-01-02-02.
This suggests that no statistically significant QPO structures are present within the current uncertainties.

\section{Discussion} \label{sec:disc}

In this work, we analyzed low-frequency type-C QPOs detected in GX~339--4 during the 2007 and 2010 outbursts with \textit{RXTE}/PCA.
In particular, we focused on QPOs with a central frequency ${\lesssim}\qty{1}{Hz}$ and without significant harmonics. 
Those with harmonics will be reported separately (Qin et al.\ in prep.). 
We have shown that in this QPO sample, the QPOs and BBN share similar behaviors: 
(1)~their IMFs show statistically consistent correlations in the $\Gamma$--count rate plane, and
(2)~their PSDs display statistically consistent energy dependence around the QPO frequencies.
The positive $\Gamma$--rate correlation can be explained by more efficient Compton cooling at phases with higher rate.
These results suggest that these type-C QPOs and BBN may be driven by the same physical processes in GX~339--4 --- specifically, QPOs may arise from the amplification of particular oscillation modes within the BBN \citep[e.g.,][]{2010MNRAS.404..738C,2025arXiv251209026L}.
Such a scenario can naturally explain that $\Gamma$ measured at different QPO and BBN IMF phases shows the same dependence on count rate, and the featureless PSD ratios around the QPO frequencies.
This is also in line with the speculations in previous studies, e.g., based on the same correlation between the BBN break frequency and QPO central frequency in different types of sources \citep{1999ApJ...514..939W}.

\begin{figure*}
\centering
\includegraphics[width=.48\textwidth]{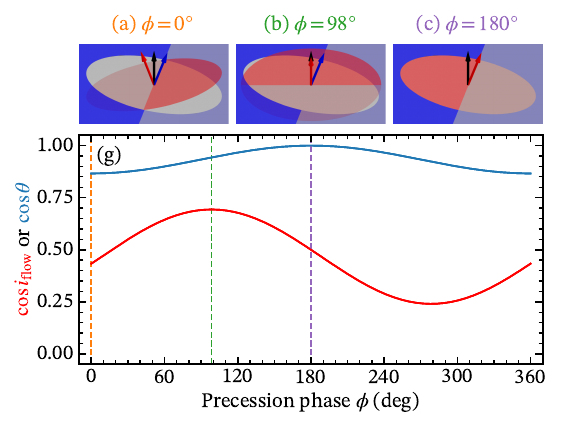}
\includegraphics[width=.48\textwidth]{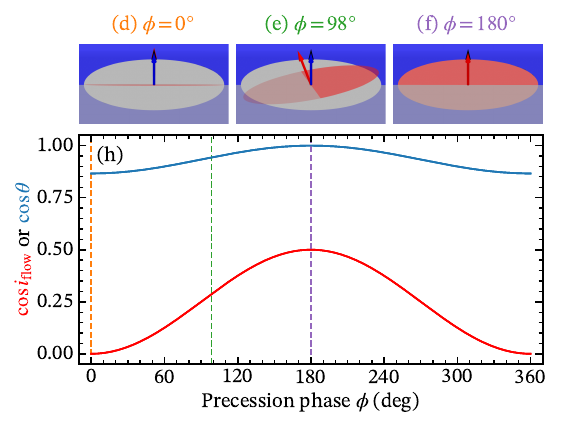}
\caption{%
Angle variations as a function of precession phases. 
$i\subrm{flow}$ is the inclination of the inner flow with respect to the line of sight. 
$\theta$ is the angle between the inner flow and outer disk. 
The black arrow marks the black hole spin axis. The blue and red arrows mark the angular momentum of the outer disk and inner flow, respectively. 
The left panel represents the general case that the outer disk momentum and black hole spin are not projected at the same direction on the sky plane, while the right panel represents the special case that they are aligned on the sky plane. 
The geometries at three different precessing phases are illustrated and the corresponding phases are marked as dashed lines.
Panels~(a--f) are generated with GeoGebra \copyright\ (\url{https://www.geogebra.org}).}
\label{fig:lt-modulation}
\end{figure*}

One of the popular models of low-frequency QPOs is that they are caused by the Lense--Thirring precession of the hot inner flow \citep{2009MNRAS.397L.101I}.
Within the same framework, \citet{2011MNRAS.415.2323I} proposed physical mechanisms for both QPOs and BBN: 
while the Lense--Thirring precession generates QPOs through geometric effects, BBN arises from intrinsic fluctuations in the accretion rate.
Via simulations, \citet{2018ApJ...858...82Y} found that the flux peaks when the hot inner flow is viewed face-on (e.g., \autoref{fig:lt-modulation}(b) and (f)), while $\Gamma$ peaks when the inner flow aligns with the outer disk.
This suggests that, in general, the peak flux\footnote{Since the spectral variation is small, the flux and count rate are effectively equivalent.} and maximum $\Gamma$ are not aligned during precessing, unless the spin axis of the black hole and the angular momentum of the outer disk are projected at the same direction on the sky plane. 
We illustrate these in \autoref{fig:lt-modulation}.
As one can see, a phase lag between the modulations of flux and $\Gamma$ is expected, unless the system is viewed at a special angle.
Instead, a precessing oblate corona may produce an in-phase modulation between the QPO flux and $\Gamma$ \citep{2023ApJ...943..165S,2023ApJ...957...84S}.
For a precessing accelerating jet, high energy photons originate from low speed blobs are more isotropic than low energy photons from high speed blobs, leading to a softer spectrum at high flux phases \citep{2023ApJ...957...84S}, also consistent with what we see in \autoref{fig:phase-resolved-results}(b).
However, the $\Gamma$ modulation in these models is due to geometric effects, and thus one would expect a different $\Gamma$--rate relation from that of BBN. 
It would be coincidental to explain the same relation presented in \autoref{fig:phase-resolved-results}(b).

It is also difficult for the precession model to explain the featureless PSD ratios around the QPO frequencies seen in \autoref{fig:psd-ratio}.
In the precession model for QPOs, the predicted rms--energy spectrum of the QPOs depends on various factors, e.g., viewing directions, geometry of the hot inner flow, black hole spin and truncation radius of the outer disk \citep{2018ApJ...858...82Y}.
The energy dependence of the BBN power density is related to the fluctuation propagation process.
It is difficult for the very different mechanisms to produce exactly the same energy dependences of the power density.
Since the power of QPOs dominates at the QPO central frequency, and the BBN power dominates away from the QPO, the precession model would predict an abrupt ratio change around the QPO frequencies.

Geometric models for QPOs were favored in some observational findings. 
For example, the type-C QPO rms and phase lags have been found to be different between systems of low and high inclinations \citep{2015MNRAS.447.2059M, 2015MNRAS.448.3348H, 2017MNRAS.464.2643V}.
As we mentioned above, we focused on QPOs below $\qty{1}{Hz}$ and without significant harmonics. 
As displayed in Figure~1 of \citet{2015MNRAS.447.2059M} and Figure~3 of \citet{2017MNRAS.464.2643V}, both the rms and phase lags of such QPOs are consistent between systems of different inclinations.
We have found that QPOs at ${\gtrsim}\qty{1}{Hz}$ and with harmonics exhibit distinct properties, and will report the results in a subsequent work (Qin et al.\ in prep.).
Furthermore, BBN was also found to display statistically different rms and phase lags across sources of different inclinations \citep{2015MNRAS.447.2059M, 2017MNRAS.464.2643V}, casting doubt on its causal link with geometric effects.
Moreover, the lack of X-ray polarization modulation over QPO phases in Swift~J1727.8--1613 appears to disfavor a geometric origin of QPOs, unless the system is viewed nearly edge-on \citep{2024ApJ...961L..42Z}.

On the other hand, several studies have proposed that QPOs are in fact due to instabilities and/or oscillations of the accretion process itself. 
\citet{2010MNRAS.404..738C} suggested that, the externally excited perturbations travel across the corona and reflect at the inner radius.
The waves satisfying resonant conditions are amplified as QPOs; otherwise, they manifest as BBN.
In this model, the BBN and QPOs result from the same physical process and should behave the same spectroscopically.
\citet{2011ApJ...736..107O} revealed quasi-periodic azimuthal magnetic field reversals due to magnetohydrodynamic dynamo and linked them to QPOs.
Since no radiative processes are included in their simulations, it is hard to predict the detailed spectral responses.
Nevertheless, since accretion is also a magnetohydrodynamic process \citep{1991ApJ...376..214B}, it would be possible that BBN and QPOs behave the same spectroscopically in this model.
Recently, \citet{2025arXiv251209026L} proposed a corona self-oscillation model for QPOs and predicted softer spectra (lower corona temperature) at higher QPO fluxes driven by a cycle of pair production and Compton cooling, in excellent agreement with what we see in \autoref{fig:phase-resolved-results}(b). 
A more detailed formulation for their model is needed to make further comparisons with data.
This cooler--brighter relation ($\Gamma$ positively correlates with count rate) could be a general feature of other QPO models based on coronal oscillation, as it reflects the fundamental nature of a corona that cools as the seed photon flux or optical depth increases.

The rms--energy relations for both QPOs and BBN have been intensively studied in the literature \citep[e.g.,][]{1997A&A...322..857B, 1999ApJS..124..265R, 2010ApJ...714.1065R, 2017ApJ...845..143Z}.
These studies fit the PSDs in different energy bands and derived the integrated rms of the QPO or BBN components.
Distinct correlation patterns have been seen during an outburst of GX~339--4 \citep{2011BASI...39..409B,2017ApJ...845..143Z}, while the relations for QPOs and BBN can be different \citep{1999ApJS..124..265R}.
Our results suggest that the BBN component around the QPO frequency has the same rms--energy relation with QPOs --- specifically, no statistically significant QPO-like structures are detected in the ratio spectra (\autoref{fig:psd-ratio}).
The ratios at other frequencies are not constant, indicating that our results do not contradict with previous ones.
Therefore, these results further favor the scenario that QPOs are excited BBN at a narrow frequency band.

We again emphasize that our conclusions are derived from the GX~339--4 type-C QPOs in the rising phase of the 2007 and 2010 outbursts without significant harmonics.
The results from those with significant harmonics will be presented in a subsequent work (Qin et al.\ in prep.).
Further systematic studies are needed to extend these conclusions to other BHXRBs.

\section{Conclusions}
\label{sec:conclusion}

In this work, we improved the techniques for QPO phase determination. Key technical results are summarized below.
\begin{itemize}[nosep]
    \item We fixed the QPO frequency and width in the VMD with values obtained from PSD fitting, allowing for a robust IMF decomposition.
    \item The IMF in the white noise region also displays spectral modulation due to background variation over phases; this can be resolved by using independent instruments for phase determination and spectral analysis, respectively.  
    \item IMFs under the BBN component can be used to study the BBN behavior in a narrow band. 
\end{itemize}

We applied the techniques to the analysis of low-frequency type-C QPOs, specifically those ${\lesssim}\qty{1}{Hz}$ and without significant harmonics, from GX~339--4 during the 2007 and 2010 outbursts, and obtained the following scientific results:
\begin{itemize}[nosep]
    \item The IMFs constructed at the QPOs and around the BBN show statistically consistent behaviors: $\Gamma$ is positively correlated with count rate as a function of phase.
    \item There are no statistically significant signatures of QPOs in the ratio spectra between PSDs at different energies.
    \item These suggest that QPOs and BBN may be driven by the same physical process.
    \item The QPO models due to geometric modulation are difficult to explain the results, while those based on corona oscillations are favored.
\end{itemize}

\begin{acknowledgments}
We thank the anonymous referee for useful comments.
HF acknowledges funding support from the National Natural Science Foundation of China under the grant 12025301, the Strategic Priority Research Program of the Chinese Academy of Sciences, and China's Space Origins Exploration Program.
JQ thanks Mingyu Ge for the help with the issue about \textit{RXTE}/PCA response generation.
This research has made use of data and software provided by the High Energy Astrophysics Science Archive Research Center (HEASARC), which is a service of the Astrophysics Science Division at NASA/GSFC.
\end{acknowledgments}

\facilities{RXTE (PCA)}

\software{%
  \texttt{Astropy} \citep{2013A&A...558A..33A,2018AJ....156..123A,2022ApJ...935..167A},
  \texttt{Stingray} \citep{2019ApJ...881...39H,2024JOSS....9.7389B,2024zndo..13974481B},
  \texttt{PyXspec} \citep{2021ascl.soft01014G}
}

\appendix

\section{Spurious Modulations from Poisson Noise} \label{appx:white-noise-modulation}

\begin{figure}
\centering
\includegraphics[width=.95\textwidth]{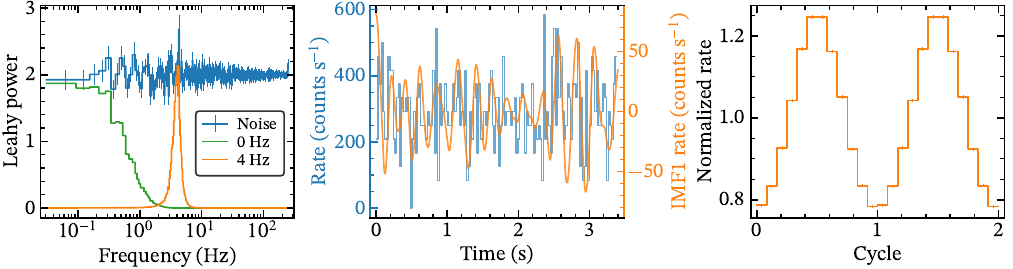}
\caption{%
Simulations of spurious modulation extracted from a white noise IMF. 
Left:~PSD and IMFs. IMF0 is the zero-frequency IMF and IMF1 is the one of interest.
Middle:~a segment of simulated light curve (blue) and the extracted IMF1 waveform (orange).
Right:~folded count rate profile at different phases normalized to the mean rate.}
\label{fig:white-noise-sim}
\end{figure}

With fixed frequency and bandwidth, our improved VMD technique can also be applied onto white noise dominated regions where no QPOs present. 
As shown in \autoref{fig:white-noise-sim}, the IMF tries to fit the Poisson fluctuations, producing spurious count rate modulations over pseudo-phases.

\begin{figure*}
\centering
\includegraphics[width=.95\textwidth]{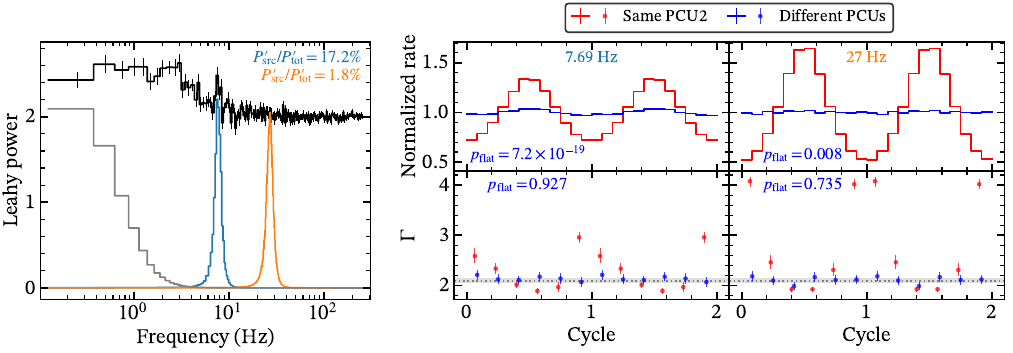}
\caption{Spurious modulations extracted from a white noise IMF from a PCA observation of GX~339--4 (ObsID = 60705-01-56-00, not used for scientific analysis). 
Left:~PSD in $\qtyrange{2.1}{20.6}{keV}$, which contains weak type-C QPOs around $\qty{7.69}{Hz}$. Three IMFs are imposed: one at zero frequency, one on the QPOs, and one on the Poisson noise ($\qty{27}{Hz}$). 
The fractional power of non--white noise components for the QPO and white noise IMF is shown in the top right corner, respectively.
Right:~normalized count rate and photon index variations as a function of phase extracted from the two nonzero-frequency IMFs. Red curves and points are produced using PCU2 data alone. Blue curves and points are produced using PCUs 0 and 3 for phase determination and PCU2 for spectral analysis. The dashed horizontal line and shaded area represent $\Gamma$ and its error range by fitting the time-averaged spectrum in $\qtyrange{2}{30}{keV}$.
The $p$-values under a constant hypothesis for the data obtained from different PCUs are shown in the panel.}
\label{fig:white-noise-obs}
\end{figure*}

\begin{figure}
\centering
\includegraphics[width=.5\columnwidth]{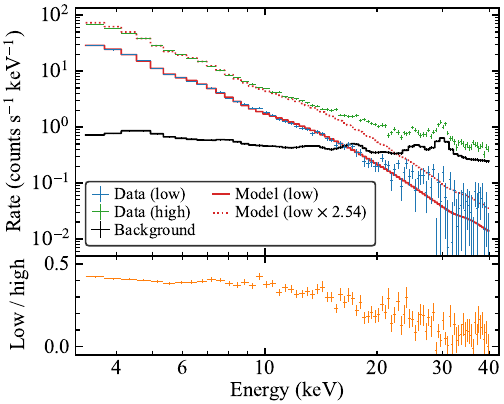}
\caption{Background subtracted spectra of ObsID 60705-01-56-00. 
The low-rate spectrum is extracted from $\qty[parse-numbers=false]{2^{-9}}{s}$ time bins containing only one count, while the high-rate spectrum is extracted from bins with two counts or more. The estimated background spectrum is shown in black. 
The red solid line represents the best-fit model to the low-rate spectrum, and the red dotted line shows the same model scaled up by a factor of 2.54.
The lower panel shows the ratio between the low- and high-rate spectra.}
\label{fig:spec-high-low-rate}
\end{figure}

For white noise, one still expects phase-independent spectroscopy due to independent fluctuations at different phases. 
However, as demonstrated in \autoref{fig:white-noise-obs}, when the technique is applied to a real observation, significant count rate and spectral modulations are seen from an IMF around $\qty{27}{Hz}$, where the white noise dominates.
This is because for non-imaging instruments like PCA, the estimate of background is model dependent, and the same model/parameters are used to estimate the background at different phases.
Even for an imaging device for which the mean background can be subtracted, the Poisson fluctuations remain in the data.
As a consequence, at phases where the fluctuations are higher than average, the background is underestimated, leading to a biased spectral shape (harder or smaller $\Gamma$), and vice versa (see \autoref{fig:spec-high-low-rate}).

Such an effect is not important around strong QPOs, because the power of QPOs is much stronger than that of white noise. 
To completely eliminate the effect, we always used PCU2 data to extract spectra, and used the light curves summed over other PCU(s) to calculate the phases via VMD, because the white noise is not correlated between detectors. 
As shown in \autoref{fig:white-noise-obs}, spurious count rate or spectral modulations are significantly depressed.
Another approach is to scale the estimated background using the source rate in each phase bin.
This may eliminate the spurious spectral modulation if the background spectrum is phase independent, but cannot remove the spurious count rate modulation.

\figsetstart
\figsetnum{1}
\figsettitle{Phase-resolved photon index modulations of all observations used in this work.}

\figsetgrpstart
\figsetgrpnum{1.1}
\figsetgrptitle{92035-01-01-01}
\figsetplot{phase-resolved-results-2007_v5.pdf[1]} 
\figsetgrpnote{Phase-resolved photon index modulations of observation 92035-01-01-01.}
\figsetgrpend

\figsetgrpstart
\figsetgrpnum{1.2}
\figsetgrptitle{92035-01-01-03}
\figsetplot{phase-resolved-results-2007_v5.pdf[2]} 
\figsetgrpnote{Phase-resolved photon index modulations of observation 92035-01-01-03.}
\figsetgrpend

\figsetgrpstart
\figsetgrpnum{1.3}
\figsetgrptitle{92035-01-01-02}
\figsetplot{phase-resolved-results-2007_v5.pdf[3]} 
\figsetgrpnote{Phase-resolved photon index modulations of observation 92035-01-01-02.}
\figsetgrpend

\figsetgrpstart
\figsetgrpnum{1.4}
\figsetgrptitle{92035-01-01-04}
\figsetplot{phase-resolved-results-2007_v5.pdf[4]} 
\figsetgrpnote{Phase-resolved photon index modulations of observation 92035-01-01-04.}
\figsetgrpend

\figsetgrpstart
\figsetgrpnum{1.5}
\figsetgrptitle{92035-01-02-00}
\figsetplot{phase-resolved-results-2007_v5.pdf[5]} 
\figsetgrpnote{Phase-resolved photon index modulations of observation 92035-01-02-00.}
\figsetgrpend

\figsetgrpstart
\figsetgrpnum{1.6}
\figsetgrptitle{92035-01-02-01}
\figsetplot{phase-resolved-results-2007_v5.pdf[6]} 
\figsetgrpnote{Phase-resolved photon index modulations of observation 92035-01-02-01.}
\figsetgrpend

\figsetgrpstart
\figsetgrpnum{1.7}
\figsetgrptitle{92035-01-02-02}
\figsetplot{phase-resolved-results-2007_v5.pdf[7]} 
\figsetgrpnote{Phase-resolved photon index modulations of observation 92035-01-02-02.}
\figsetgrpend

\figsetgrpstart
\figsetgrpnum{1.8}
\figsetgrptitle{92035-01-02-03}
\figsetplot{phase-resolved-results-2007_v5.pdf[8]} 
\figsetgrpnote{Phase-resolved photon index modulations of observation 92035-01-02-03.}
\figsetgrpend

\figsetgrpstart
\figsetgrpnum{1.9}
\figsetgrptitle{92035-01-02-04}
\figsetplot{phase-resolved-results-2007_v5.pdf[9]} 
\figsetgrpnote{Phase-resolved photon index modulations of observation 92035-01-02-04.}
\figsetgrpend

\figsetgrpstart
\figsetgrpnum{1.10}
\figsetgrptitle{92035-01-02-08}
\figsetplot{phase-resolved-results-2007_v5.pdf[10]} 
\figsetgrpnote{Phase-resolved photon index modulations of observation 92035-01-02-08.}
\figsetgrpend

\figsetgrpstart
\figsetgrpnum{1.11}
\figsetgrptitle{92035-01-02-07}
\figsetplot{phase-resolved-results-2007_v5.pdf[11]} 
\figsetgrpnote{Phase-resolved photon index modulations of observation 92035-01-02-07.}
\figsetgrpend

\figsetgrpstart
\figsetgrpnum{1.12}
\figsetgrptitle{95409-01-12-01}
\figsetplot{phase-resolved-results-2010_v5.pdf[1]} 
\figsetgrpnote{Phase-resolved photon index modulations of observation 95409-01-12-01.}
\figsetgrpend

\figsetgrpstart
\figsetgrpnum{1.13}
\figsetgrptitle{95409-01-13-03}
\figsetplot{phase-resolved-results-2010_v5.pdf[2]} 
\figsetgrpnote{Phase-resolved photon index modulations of observation 95409-01-13-03.}
\figsetgrpend

\figsetgrpstart
\figsetgrpnum{1.14}
\figsetgrptitle{95409-01-13-04}
\figsetplot{phase-resolved-results-2010_v5.pdf[3]} 
\figsetgrpnote{Phase-resolved photon index modulations of observation 95409-01-13-04.}
\figsetgrpend

\figsetgrpstart
\figsetgrpnum{1.15}
\figsetgrptitle{95409-01-13-02}
\figsetplot{phase-resolved-results-2010_v5.pdf[4]} 
\figsetgrpnote{Phase-resolved photon index modulations of observation 95409-01-13-02.}
\figsetgrpend

\figsetgrpstart
\figsetgrpnum{1.16}
\figsetgrptitle{95409-01-13-05}
\figsetplot{phase-resolved-results-2010_v5.pdf[5]} 
\figsetgrpnote{Phase-resolved photon index modulations of observation 95409-01-13-05.}
\figsetgrpend

\figsetgrpstart
\figsetgrpnum{1.17}
\figsetgrptitle{95409-01-13-06}
\figsetplot{phase-resolved-results-2010_v5.pdf[6]} 
\figsetgrpnote{Phase-resolved photon index modulations of observation 95409-01-13-06.}
\figsetgrpend

\figsetgrpstart
\figsetgrpnum{1.18}
\figsetgrptitle{95409-01-14-00}
\figsetplot{phase-resolved-results-2010_v5.pdf[7]} 
\figsetgrpnote{Phase-resolved photon index modulations of observation 95409-01-14-00.}
\figsetgrpend

\figsetgrpstart
\figsetgrpnum{1.19}
\figsetgrptitle{95409-01-14-01}
\figsetplot{phase-resolved-results-2010_v5.pdf[8]} 
\figsetgrpnote{Phase-resolved photon index modulations of observation 95409-01-14-01.}
\figsetgrpend

\figsetend

\figsetstart
\figsetnum{2}
\figsettitle{PSDs in four energy bands and their ratios against the lowest one, for all observations used in this work.}

\figsetgrpstart
\figsetgrpnum{2.1}
\figsetgrptitle{92035-01-01-01}
\figsetplot{psd-ratio-2007_v2.pdf[1]} 
\figsetgrpnote{Multi-band PSDs and their ratios of observation 92035-01-01-01.}
\figsetgrpend

\figsetgrpstart
\figsetgrpnum{2.2}
\figsetgrptitle{92035-01-01-03}
\figsetplot{psd-ratio-2007_v2.pdf[2]} 
\figsetgrpnote{Multi-band PSDs and their ratios of observation 92035-01-01-03.}
\figsetgrpend

\figsetgrpstart
\figsetgrpnum{2.3}
\figsetgrptitle{92035-01-01-02}
\figsetplot{psd-ratio-2007_v2.pdf[3]} 
\figsetgrpnote{Multi-band PSDs and their ratios of observation 92035-01-01-02.}
\figsetgrpend

\figsetgrpstart
\figsetgrpnum{2.4}
\figsetgrptitle{92035-01-01-04}
\figsetplot{psd-ratio-2007_v2.pdf[4]} 
\figsetgrpnote{Multi-band PSDs and their ratios of observation 92035-01-01-04.}
\figsetgrpend

\figsetgrpstart
\figsetgrpnum{2.5}
\figsetgrptitle{92035-01-02-00}
\figsetplot{psd-ratio-2007_v2.pdf[5]} 
\figsetgrpnote{Multi-band PSDs and their ratios of observation 92035-01-02-00.}
\figsetgrpend

\figsetgrpstart
\figsetgrpnum{2.6}
\figsetgrptitle{92035-01-02-01}
\figsetplot{psd-ratio-2007_v2.pdf[6]} 
\figsetgrpnote{Multi-band PSDs and their ratios of observation 92035-01-02-01.}
\figsetgrpend

\figsetgrpstart
\figsetgrpnum{2.7}
\figsetgrptitle{92035-01-02-02}
\figsetplot{psd-ratio-2007_v2.pdf[7]} 
\figsetgrpnote{Multi-band PSDs and their ratios of observation 92035-01-02-02.}
\figsetgrpend

\figsetgrpstart
\figsetgrpnum{2.8}
\figsetgrptitle{92035-01-02-03}
\figsetplot{psd-ratio-2007_v2.pdf[8]} 
\figsetgrpnote{Multi-band PSDs and their ratios of observation 92035-01-02-03.}
\figsetgrpend

\figsetgrpstart
\figsetgrpnum{2.9}
\figsetgrptitle{92035-01-02-04}
\figsetplot{psd-ratio-2007_v2.pdf[9]} 
\figsetgrpnote{Multi-band PSDs and their ratios of observation 92035-01-02-04.}
\figsetgrpend

\figsetgrpstart
\figsetgrpnum{2.10}
\figsetgrptitle{92035-01-02-08}
\figsetplot{psd-ratio-2007_v2.pdf[10]} 
\figsetgrpnote{Multi-band PSDs and their ratios of observation 92035-01-02-08.}
\figsetgrpend

\figsetgrpstart
\figsetgrpnum{2.11}
\figsetgrptitle{92035-01-02-07}
\figsetplot{psd-ratio-2007_v2.pdf[11]} 
\figsetgrpnote{Multi-band PSDs and their ratios of observation 92035-01-02-07.}
\figsetgrpend

\figsetgrpstart
\figsetgrpnum{2.12}
\figsetgrptitle{95409-01-12-01}
\figsetplot{psd-ratio-2010_v2.pdf[1]} 
\figsetgrpnote{Multi-band PSDs and their ratios of observation 95409-01-12-01.}
\figsetgrpend

\figsetgrpstart
\figsetgrpnum{2.13}
\figsetgrptitle{95409-01-13-03}
\figsetplot{psd-ratio-2010_v2.pdf[2]} 
\figsetgrpnote{Multi-band PSDs and their ratios of observation 95409-01-13-03.}
\figsetgrpend

\figsetgrpstart
\figsetgrpnum{2.14}
\figsetgrptitle{95409-01-13-04}
\figsetplot{psd-ratio-2010_v2.pdf[3]} 
\figsetgrpnote{Multi-band PSDs and their ratios of observation 95409-01-13-04.}
\figsetgrpend

\figsetgrpstart
\figsetgrpnum{2.15}
\figsetgrptitle{95409-01-13-02}
\figsetplot{psd-ratio-2010_v2.pdf[4]} 
\figsetgrpnote{Multi-band PSDs and their ratios of observation 95409-01-13-02.}
\figsetgrpend

\figsetgrpstart
\figsetgrpnum{2.16}
\figsetgrptitle{95409-01-13-05}
\figsetplot{psd-ratio-2010_v2.pdf[5]} 
\figsetgrpnote{Multi-band PSDs and their ratios of observation 95409-01-13-05.}
\figsetgrpend

\figsetgrpstart
\figsetgrpnum{2.17}
\figsetgrptitle{95409-01-13-06}
\figsetplot{psd-ratio-2010_v2.pdf[6]} 
\figsetgrpnote{Multi-band PSDs and their ratios of observation 95409-01-13-06.}
\figsetgrpend

\figsetgrpstart
\figsetgrpnum{2.18}
\figsetgrptitle{95409-01-14-00}
\figsetplot{psd-ratio-2010_v2.pdf[7]} 
\figsetgrpnote{Multi-band PSDs and their ratios of observation 95409-01-14-00.}
\figsetgrpend

\figsetgrpstart
\figsetgrpnum{2.19}
\figsetgrptitle{95409-01-14-01}
\figsetplot{psd-ratio-2010_v2.pdf[8]} 
\figsetgrpnote{Multi-band PSDs and their ratios of observation 95409-01-14-01.}
\figsetgrpend

\figsetend

\bibliography{refs}{}
\bibliographystyle{aasjournalv7.1}

\end{CJK*}
\end{document}